\documentclass[aps,prl,reprint,twocolumn,superscriptaddress]{revtex4-1}

\usepackage{amsthm}
\usepackage{amsmath}
\usepackage{graphicx}
\usepackage{slashed}
\usepackage{amssymb}
\usepackage{float}
\usepackage[colorlinks=True, citecolor=blue, urlcolor=blue, linkcolor=blue]{hyperref}
\usepackage[braket,qm]{qcircuit}

\newcommand{\nn}{\nonumber}

\newcommand{\be}{\begin{eqnarray}}
\newcommand{\ee}{\end{eqnarray}}

\begin{document}

\title{The soft drop momentum sharing fraction $z_g$ beyond leading-logarithmic accuracy}

\author{Pedro Cal}
\email{p.cal@nikhef.nl}
\affiliation{Institute for Theoretical Physics Amsterdam and Delta Institute for Theoretical Physics, University of Amsterdam, Science Park 904, 1098 XH Amsterdam, The Netherlands}
\affiliation{Nikhef, Theory Group, Science Park 105, 1098 XG, Amsterdam, The Netherlands}

\author{Kyle Lee}
\email{kylelee@lbl.gov}
\affiliation{Department of Physics, University of California, Berkeley, CA 94720, USA}
\affiliation{Nuclear Science Division, Lawrence Berkeley National Laboratory, Berkeley, California 94720, USA}

\author{Felix Ringer}
\email{fmringer@lbl.gov}
\affiliation{Nuclear Science Division, Lawrence Berkeley National Laboratory, Berkeley, California 94720, USA}

\author{Wouter J. Waalewijn}
\email{w.j.waalewijn@uva.nl}
\affiliation{Institute for Theoretical Physics Amsterdam and Delta Institute for Theoretical Physics, University of Amsterdam, Science Park 904, 1098 XH Amsterdam, The Netherlands}
\affiliation{Nikhef, Theory Group, Science Park 105, 1098 XG, Amsterdam, The Netherlands}

\date{\today}

\begin{abstract}
Grooming techniques, such as soft drop, play a central role in reducing sensitivity of jets to e.g.~underlying event and hadronization at current collider experiments. The momentum sharing fraction $z_g$, of the two branches in a jet that pass the soft drop condition, is one of the most important observables characterizing a collinear splitting inside the jet, and directly probes the QCD splitting functions. In this work, we present a factorization framework that enables a systematic calculation of the corresponding cross section beyond leading-logarithmic (LL) accuracy, showing that this measurement is not only sensitive to the QCD charge but also the spin of the parton that initiates the jet. Our results at next-to-leading logarithmic (NLL$'$) accuracy include non-global logarithms, and provide a first meaningful assessment of the perturbative uncertainty. We present a comparison to the available experimental data from ALICE, ATLAS, and STAR and find excellent agreement.
\end{abstract}

\maketitle

{\it Introduction.} At high-energy collider experiments, jets and their substructure play a central role in probing fundamental aspects of QCD and searching for physics beyond the standard model~\cite{Larkoski:2017jix,Asquith:2018igt,Marzani:2019hun}. Jet grooming techniques~\cite{Krohn:2009th,Ellis:2009me,Dasgupta:2013ihk,Cacciari:2014gra,Larkoski:2014wba} are designed to  remove soft radiation inside jets, crucially reducing contamination in the complicated environment of hadron colliders. These techniques will become even more important during the high-luminosity era of the LHC. Grooming algorithms can also lead to significantly reduced nonperturbative (hadronization) corrections, allowing for direct and precise comparisons between theory and data, see e.g.~Refs.~\cite{Aaboud:2017qwh,Sirunyan:2018xdh,Aad:2019vyi}.

\begin{figure}[t]
\includegraphics[scale=0.22]{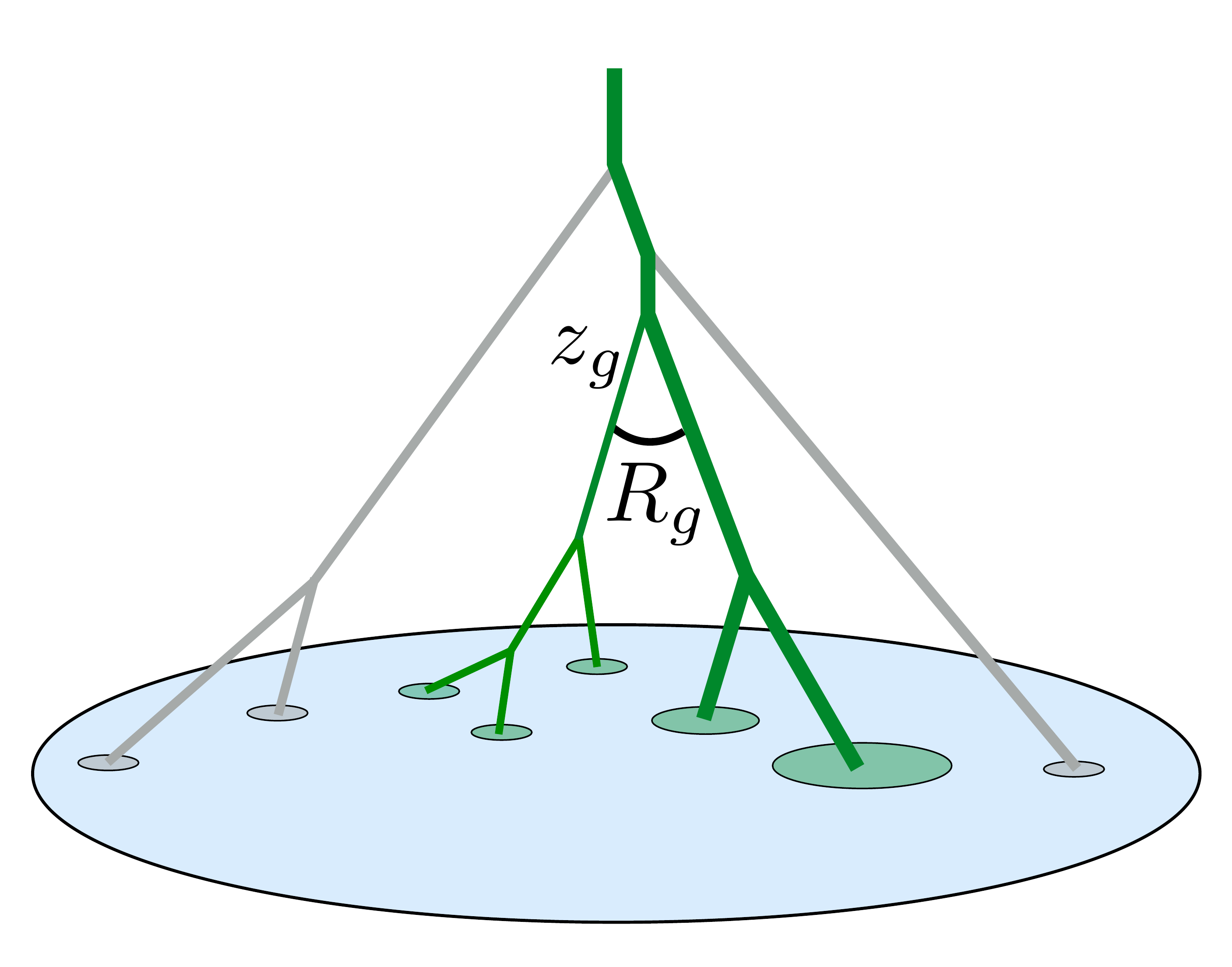}
\caption{Illustration of the clustering tree of a jet: Branches are groomed away (grey) until the soft drop condition in Eq.~(\ref{eq:SD_condt}) is first satisfied. This splitting sets the observables $z_g$ and $R_g$, and subsequent splittings (green) are kept.~\label{fig:intro}}
\end{figure}

In this letter, we study the soft drop grooming algorithm~\cite{Larkoski:2014wba}. After reclustering a jet with the Cambridge/Aachen (C/A)~\cite{Dokshitzer:1997in,Wobisch:1998wt} algorithm, it iteratively declusters the jet, at each step removing the softer branch if its momentum fraction $z$ fails the soft drop condition
\begin{equation}\label{eq:SD_condt}
z>z_{\mathrm{cut}}\left(\Delta R_{12}/R\right)^{\beta} \,.
\end{equation}
Here, $\Delta R_{12}$ is the distance between the branches in the $\eta$-$\phi$ plane, $R$ is the radius of the initial ungroomed jet, and $z_{\rm cut},\beta$ are tuneable grooming parameters. (Soft drop grooming with $\beta=0$ corresponds to the modified mass drop tagger~\cite{Dasgupta:2013ihk}.) Once Eq.~\eqref{eq:SD_condt} is satisfied, the algorithm terminates and $z_g = z$ and $R_g=\Delta R_{12}$, as illustrated in Fig.~\ref{fig:intro}. Both $z_g$ and $R_g$ are central to characterizing the two hard branches of the groomed jet. 

The momentum sharing fraction $z_g$ has received a lot of attention by both the theoretical and experimental particle and nuclear physics communities in the past years. 
The main reason is that it allows for the most direct measurement of the QCD (Altarelli-Parisi) splitting functions~\cite{Altarelli:1977zs}, providing a glimpse into fundamental splittings at parton level. 
The cross section differential in $z_g$ was measured by the ALICE~\cite{Acharya:2019djg,Mulligan:2020cnp}, ATLAS~\cite{Aad:2019vyi}, CMS~\cite{Sirunyan:2017bsd,Sirunyan:2018asm} collaborations at the LHC and by STAR~\cite{Adam:2020kug} at RHIC, which we compare to in this work. In addition, the $z_g$ distribution was also extracted from CMS open data~\cite{Larkoski:2017bvj,Tripathee:2017ybi}.

Our calculation reveals that this measurement not only probes the color charge but also the spin of the parton initiating the jet, and our precision is essential to have sensitivity to this effect in interpreting experimental results.  The momentum sharing fraction is also of great interest for heavy-ion collisions, as it probes modifications of hard-collinear splittings in the quark-gluon plasma. For recent theoretical results, see Refs.~\cite{Mehtar-Tani:2016aco,Chien:2016led,Chang:2017gkt,Li:2017wwc,Milhano:2017nzm,KunnawalkamElayavalli:2017hxo,Casalderrey-Solana:2019ubu,Caucal:2019uvr}. We also expect $z_g$ to be of great phenomenological importance at the future Electron-Ion Collider (EIC)~\cite{AbdulKhalek:2021gbh}.

The observable $z_g$ was first introduced in Ref.~\cite{Larkoski:2015lea}, where a calculation of the corresponding cross section at leading-logarithmic (LL) accuracy was performed. It was found that $z_g$ is Infrared-Collinear (IRC) safe only for $\beta<0$. For $\beta\geq 0$, $z_g$ is IRC unsafe but calculable: the IRC divergence is tamed by accounting for the Sudakov suppression, making $z_g$ a Sudakov safe observable~\cite{Larkoski:2013paa}. The cross section can thus be calculated by performing the joint resummation of the logarithms of $z_g$ and the groomed radius $R_g$. Alternatively, one can impose a cut on $R_g$, but this case also requires resummation. Here we extend the work of Ref.~\cite{Larkoski:2015lea} by setting up a factorization framework within Soft Collinear Effective Theory (SCET)~\cite{Bauer:2000ew,Bauer:2000yr,Bauer:2001yt, Bauer:2002nz, Beneke:2002ph}, which allows for the systematic extension beyond LL. We obtain results at next-to-leading logarithmic (NLL$'$) accuracy, accounting for non-global logarithms (NGLs)~\cite{Dasgupta:2001sh}. Our work also provides the first meaningful assessment of perturbative uncertainties, and opens the door to future calculations at higher perturbative accuracy and for a systematic treatment of nonperturbative effects~\cite{Hoang:2019ceu,Pathak:2020iue}. Theoretical calculations of other groomed observables, such as the groomed jet mass, can be found in Refs.~\cite{Frye:2016aiz,Marzani:2017mva,Larkoski:2017cqq,Kang:2018jwa,Kardos:2018kth,Lee:2019lge,Chien:2019osu,Larkoski:2020wgx,Kardos:2020gty,Anderle:2020mxj,Mehtar-Tani:2020oux,Baron:2020xoi,Makris:2021drz,Caucal:2021bae,Cal:2019gxa,Cal:2020flh,Caletti:2021oor}.

\begin{figure}[t]
  \includegraphics[width = 0.37 \textwidth]{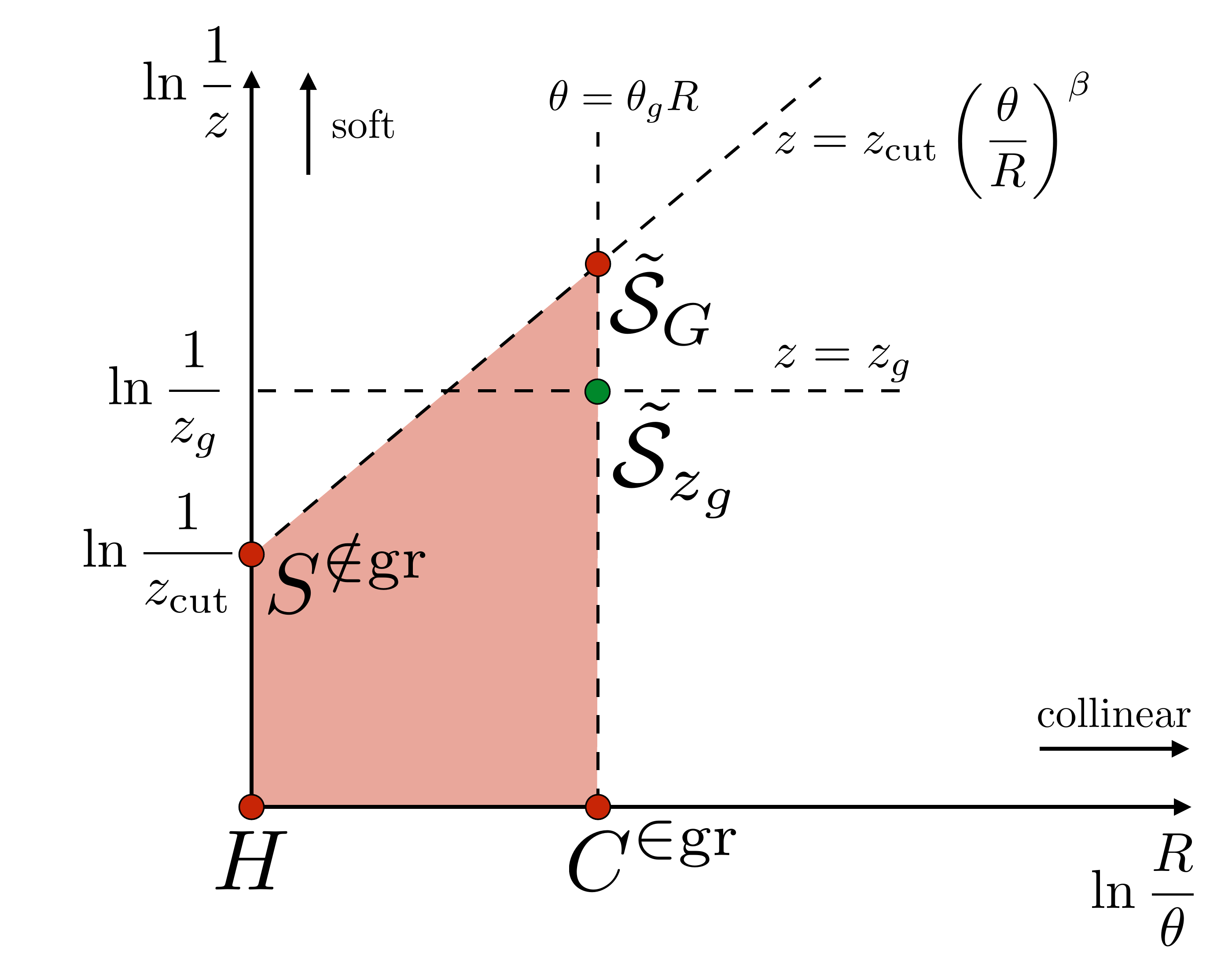}
  \caption{Lund diagram for the cross section differential in $z_g$ and $\theta_g$. The green dot corresponds to the emission passing the soft drop condition, and emissions in the red area are vetoed. The modes that appear in the corresponding SCET calculation are indicated by green and red dots.~\label{fig:Lund}}
\end{figure}

{\it Theoretical framework.} We now describe how we calculate the cross section differential in the jet's transverse momentum $p_T$ and rapidity $\eta$, as well as the groomed jet substructure observables $z_g$ and $\theta_g \equiv R_g/R$. 
For small jet radii, we can separate the production of the inclusive jet sample from the jet substructure measurement using collinear factorization,
\begin{align}\label{eq:collinear}
    \frac{{\rm d}\sigma}{{\rm d}p_T\,{\rm d}\eta\,{\rm d}z_g\,{\rm d} \theta_g}=
    & 
    \sum_{i}f_i(p_T,\eta,R,\mu) 
    \nonumber \\ &\times
    \tilde {\cal G}_i(z_g,\theta_g,p_TR,z_{\rm cut},\beta,\mu)\,.
\end{align}
The jet production is summarized by quark/gluon fractions $f_{i=q,g}$, which account for parton distribution functions, the hard-scattering, and semi-inclusive jet functions. The jet functions $\tilde {\cal G}_i$ (using the notation of Ref.~\cite{Cal:2020flh}) encode the substructure measurement. See Refs.~\cite{Aversa:1988vb,Jager:2002xm,Mukherjee:2012uz,Dasgupta:2014yra,Kaufmann:2015hma,Kang:2016mcy,Dai:2016hzf,Cal:2020flh} for more details on this first step of the factorization.

Next, we calculate the jet function $\tilde {\cal G}_i$, which we will need to describe the region where $z_g$ is order one. We find at leading order (LO) for $z_g, \theta_g>0$
\begin{align}\label{eq:fixed_order}
    \tilde  {\cal G}^{(1)}_q
    &= 
    \Theta(1/2>z_g>z_{\rm cut}\theta_g^\beta)\,
    \Theta(\theta_g<1)
    \nn \\ & \quad
    \frac{\alpha_s}{\pi} \frac{1}{\theta_g} \bigl[P_{qq}(z_g)+P_{gq}(z_g)\bigr]\,,
\end{align}
and similarly for $\tilde  {\cal G}^{(1)}_g$.
Here $P_{ij}$ are the leading-order QCD splitting functions, showing that the measurement of $z_g$ probes these. Since $\tilde {\cal G}^{(1)}_i\sim 1/\theta_g$, and there is no lower bound on $\theta_g$ for  $\beta\geq 0$, we cannot integrate out the dependence on $\theta_g$ in this case.  The integration over $\theta_g$ is only possible after taking into account the Sudakov suppression through resummation.

To achieve this,  we perform the resummation of large logarithmic corrections of $z_g,\theta_g$ and $z_{\rm cut}$. We start with the result at LL accuracy, which provides physical intuition and helps identify the relevant modes in SCET. It is described by strongly-ordered emission of gluons in the collinear and soft limit. The Lund diagram in Fig.~\ref{fig:Lund} shows the phase space of such emissions in terms of their energy fraction $z$ and angle $\theta$, with dashed lines indicating the soft drop condition in Eq.~\eqref{eq:SD_condt} and the measurement of $\theta_g$ and $z_g$. The emission at the green dot sets $z_g$ and $\theta_g$. Emissions in the red region are not allowed, and the corresponding area enters in the Sudakov exponent:
\begin{align} \label{eq:G_FO}
    \tilde {\cal G}_i
    & = \Theta(1/2>z_g>z_{\rm cut}\theta_g^\beta) 
   \frac{2\alpha_s C_i}{\pi}\frac{1}{z_g\, \theta_g}
    \nonumber\\
    & \quad
    \times \exp\bigg(-\frac{\alpha_s C_i}{\pi}(\beta \ln^2\theta_g+2\ln z_{\rm cut}\ln\theta_g)\bigg)
    \,,
\end{align}
Here $C_{i=F,A}$ denotes the appropriate color factor for quarks and gluons, which is the only dependence on the initial parton at LL. As Eq.\ \eqref{eq:G_FO} indicates, it is now safe to integrate over $\theta_g$ due to the Sudakov suppression, and the resulting expression agrees with Eq.~(14) of Ref.~\cite{Larkoski:2015lea}. 
\begin{figure*}[t]
  \includegraphics[width =  \textwidth]{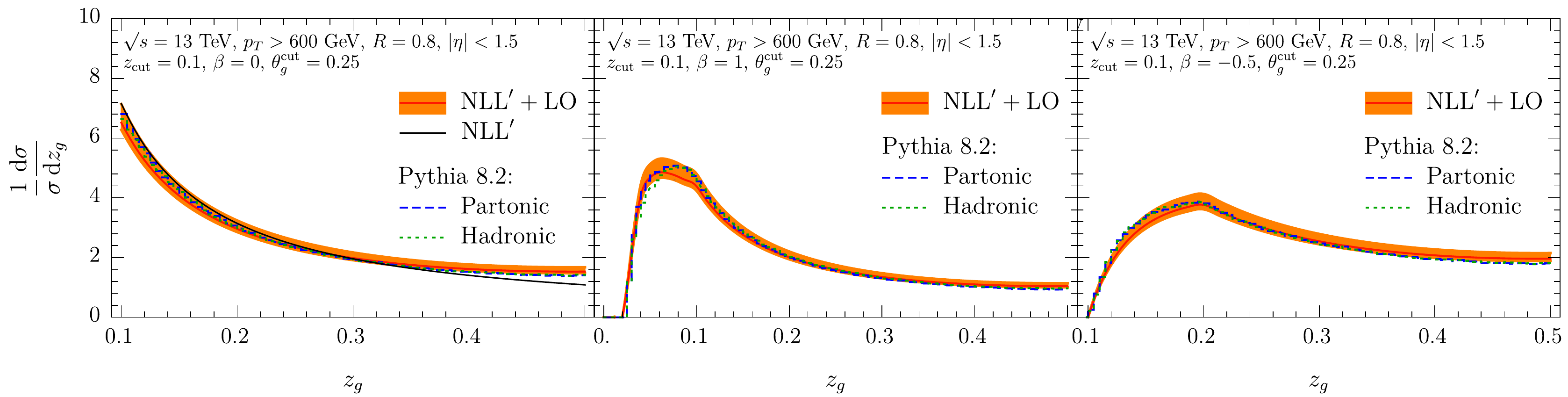}
  \caption{Comparison of our numerical results for the momentum sharing fraction $z_g$ at NLL$'$+LO to Pythia~8 simulations~\cite{Sjostrand:2014zea} at parton and hadron level, for representative jet kinematics and three choices of soft drop grooming parameters.~\label{fig:Pythia}}
\end{figure*}

We can extend this result to NLL$'$ by identifying the relevant modes within SCET, for which the scaling of the momentum components can be read off from the location of the points in the Lund diagram in Fig.~\ref{fig:Lund}.  We find
\begin{align}\label{eq:NLL}
    \tilde {\cal G}_i 
    &= \Theta(1/2>z_g>z_{\rm cut}\theta_g^\beta)    
\tilde H_i(p_TR,\mu)\,C_i^{\in {\rm gr}}(\theta_g p_T R,\mu)
    \nonumber\\& \quad
    \times S_i^{\notin {\rm gr}}(z_{\rm cut}p_T R,\beta,\mu)\,
    \tilde {\cal S}_G(z_{\rm cut}\theta_g^{1+\beta}p_T R,\beta,\mu)\    
    \nn \\ & \quad 
    \times S_i^{\rm NG}(z_{\rm cut}) \bigg[\frac{\rm d}{{\rm d}z_g} \frac{\rm d}{{\rm d}\theta_g} 
    \tilde {\cal S}_{z_g}(z_g\theta_g p_T R,\mu) 
    \nn \\ & \quad 
    + \tilde {\cal S}'^{\rm NG}_{i,1}(z_g \theta_g, z_g) 
    + \tilde {\cal S}'^{\rm NG}_{i,2}\Big(z_g \theta_g, \frac{z_g}{z_{\rm cut}\theta_g^\beta}\Big)  \bigg]\,.
\end{align}
The resummation of logarithms of $z_g, \theta_g$, and $z_{\rm cut}$ is achieved by evaluating each ingredient (except those describing NGLs) at its natural scale, which can be read off from their first argument, and using the renormalization group equations (RGEs) to evolve them to a common scale $\mu$.
The modes associated with the red points in Fig.~\ref{fig:Lund} also appeared in the factorization of the groomed jet radius~\cite{Kang:2019prh}, and the expressions for the corresponding functions can be found there. The mode corresponding to the green point is rather different because it describes the \emph{single} emission that passes the soft drop condition.
We set the $\mu$ scales for the cross section differential in $\theta_g$ and cumulative in $z_g$, and therefore present the order $\alpha_s$ expression and RGE for the new ingredient $\tilde {\cal S}_{z_g}$ differential in $\theta_g$:
\begin{align}
\frac{{\rm d}}{{\rm d} \theta_g} \tilde {\cal S}_{z_g}(z_g\theta_g p_T R,\mu)&=-\frac{2\alpha_s C_i}{\pi}\frac{1}{\theta_g} \ln\frac{\mu}{z_g \theta_g p_T R}
\,,\nn \\
\frac{{\rm d}}{{\rm d} \ln \mu}\, \frac{{\rm d}}{{\rm d} \theta_g} \tilde {\cal S}_{z_g}&= -\frac{2\alpha_s C_i}{\pi}\frac{1}{\theta_g}
\,.
\end{align}
A similarly unusual RGE was encountered in Refs.~\cite{Cal:2019gxa,Cal:2020flh}, to which we refer the reader for details.

There are three types of non-global logarithms~\cite{Dasgupta:2001sh,Banfi:2002hw,Hornig:2011iu,Kelley:2011ng,Schwartz:2014wha,Caron-Huot:2015bja,Hagiwara:2015bia,Larkoski:2015zka,Becher:2015hka,Balsiger:2018ezi,Banfi:2021owj} in Eq.~(\ref{eq:NLL}): First, the NGLs described by $S_i^{\rm NG}$ are similar to the usual hemisphere case, and arise due to correlations between the unconstrained emissions in the region outside the jet and the radiation inside the jet that fails the grooming condition. Second, the NGLs ${\cal S}_{i,1,2}'^{\rm NG}$ arise from correlated emissions, where one sets $z_g$ and $\theta_g$ (green dot) and the other is either outside the groomed radius and fails the grooming condition or inside the groomed radius and is unconstrained (corresponding to one of the two red dots on the vertical dashed line $\theta = \theta_g R$ in Fig.~\ref{fig:Lund}). These NGLs are sensitive to C/A clustering effects~\cite{Appleby:2002ke,Delenda:2006nf,Neill:2018yet}. We include their leading contribution at order $\alpha_s^2$, 
\begin{align}
    \tilde{\cal S}_{i,1}'^{{\rm NG},(2)}(z_g \theta_g, z_g) & = 2.58 C_i C_A \Big(\frac{\alpha_s}{2 \pi}\Big)^2 \frac{1}{z_g \theta_g} \ln z_g \,,
\end{align}
and the form of $\tilde{\cal S}_{i,2}'^{{\rm NG},(2)}$ is the same at this order.
In our numerical implementation these NGLs are multiplied by the global Sudakov suppression factor, see Eq.~(\ref{eq:NLL}), making their numerical size small (percent level).

{\it Numerical results and comparison to data.} Throughout this section we consider (ungroomed) jets which are reconstructed with the anti-$k_T$ algorithm~\cite{Cacciari:2008gp}, as in the measurement of the experimental collaborations. We use the parton distribution functions of Ref.~\cite{Dulat:2015mca}. 

\begin{figure}[b]
  \includegraphics[width = 0.48 \textwidth]{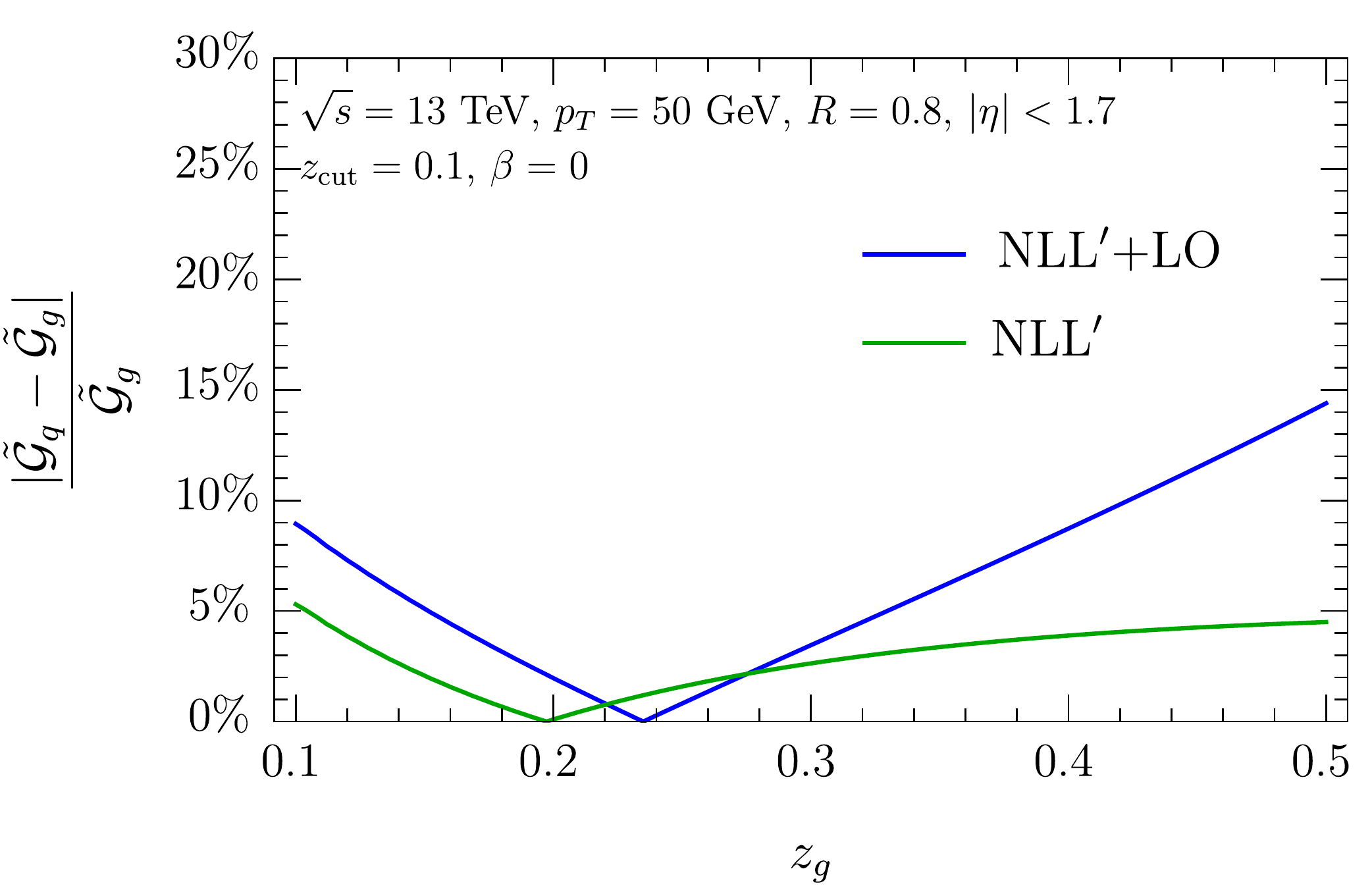}
  \caption{Relative difference between the $z_g$ distribution for quark and gluon-initiated jets. The difference between the green and blue curve indicates the size of the nonsingular part of the QCD splitting function, encoding the spin-dependence.~\label{fig:spin_color}}
\end{figure}
\begin{figure*}[t]
  \includegraphics[width = \textwidth]{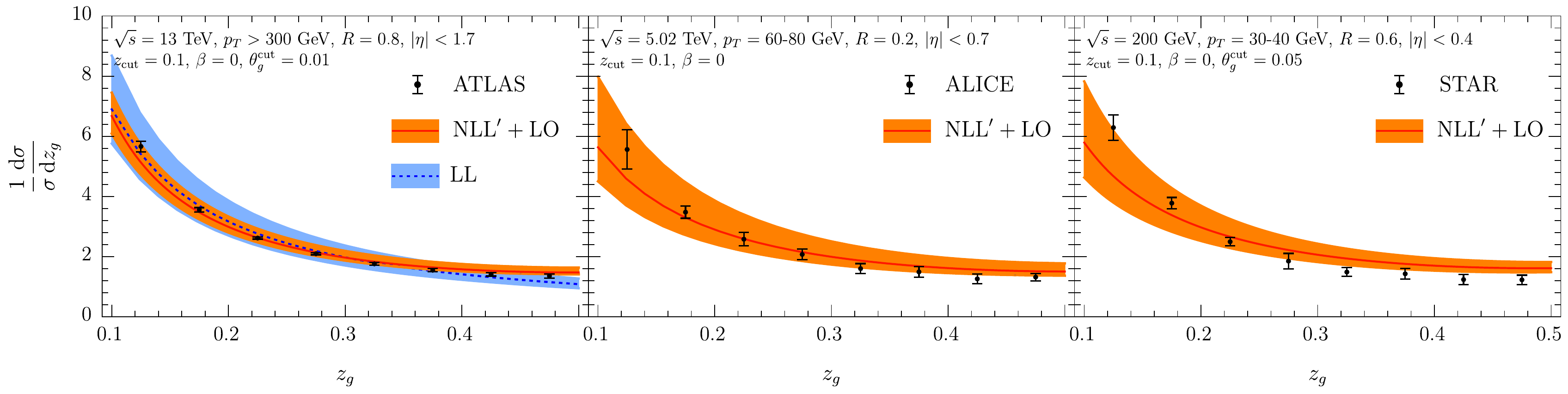}
  \caption{Comparison of our results to ATLAS~\cite{Aad:2019vyi}, ALICE~\cite{Mulligan:2020cnp} and STAR~\cite{Adam:2020kug} data for $\beta=0$.~\label{fig:data_beta0}}
\end{figure*}

We start by comparing our numerical results at NLL$'$+LO accuracy to Pythia~8 simulations~\cite{Sjostrand:2014zea}. The comparison for exemplary jet kinematics is shown in Fig.~\ref{fig:Pythia}. We choose three representative values of the grooming parameter $\beta$ and impose a cutoff on the groomed jet radius of $\theta_g^{\rm cut}>0.25$ to reduce the sensitivity to nonperturbative physics. The QCD scale uncertainty bands in the figures shown here are obtained by independently varying all relevant scales in Eqs.~\eqref{eq:collinear} and~(\ref{eq:NLL}) by a factor of 2 around the central scale choice. In addition, we smoothly freeze all scales at ${\cal O}(1~\text{GeV})$~\cite{Ligeti:2008ac}. The hadronization corrections for the chosen kinematics are very small which can be seen by comparing Pythia results at parton and hadron level. 

Overall, we observe very good agreement of our results with Pythia. In addition to the improved precision, there are notable qualitative differences with the earlier results of Ref.~\cite{Larkoski:2015lea}: The shape of our distribution for $\beta<0$ is rather different, and we find a smooth transition to $\beta=0$ for $z_g > z_{\rm cut}$, whether we approach from negative or positive $\beta$.
For large $z_g$ the matching to the LO in Eq.~(\ref{eq:fixed_order}) is essential, which is done multiplicatively because of the common singularity in $\theta_g$. Indeed, the non-singular terms added by the matching are important to achieve good agreement with Pythia, as illustrated in the left panel of Fig.~\ref{fig:Pythia} by the black NLL$'$ curve (that does not include the matching). This demonstrates the sensitivity of $z_g$ distribution to the full splitting function beyond the leading $1/z_g$ behavior in the singular limit.  
The size of these corrections, which encode the spin-dependence of the splitting function, are visualized in Fig.~\ref{fig:spin_color}. Since their size can be up to order 10\%, we expect that this can be probed experimentally by for example comparing inclusive vs.~photon-tagged jets or jets with different rapidities.

\begin{figure}[b]
  \includegraphics[width = 0.48 \textwidth]{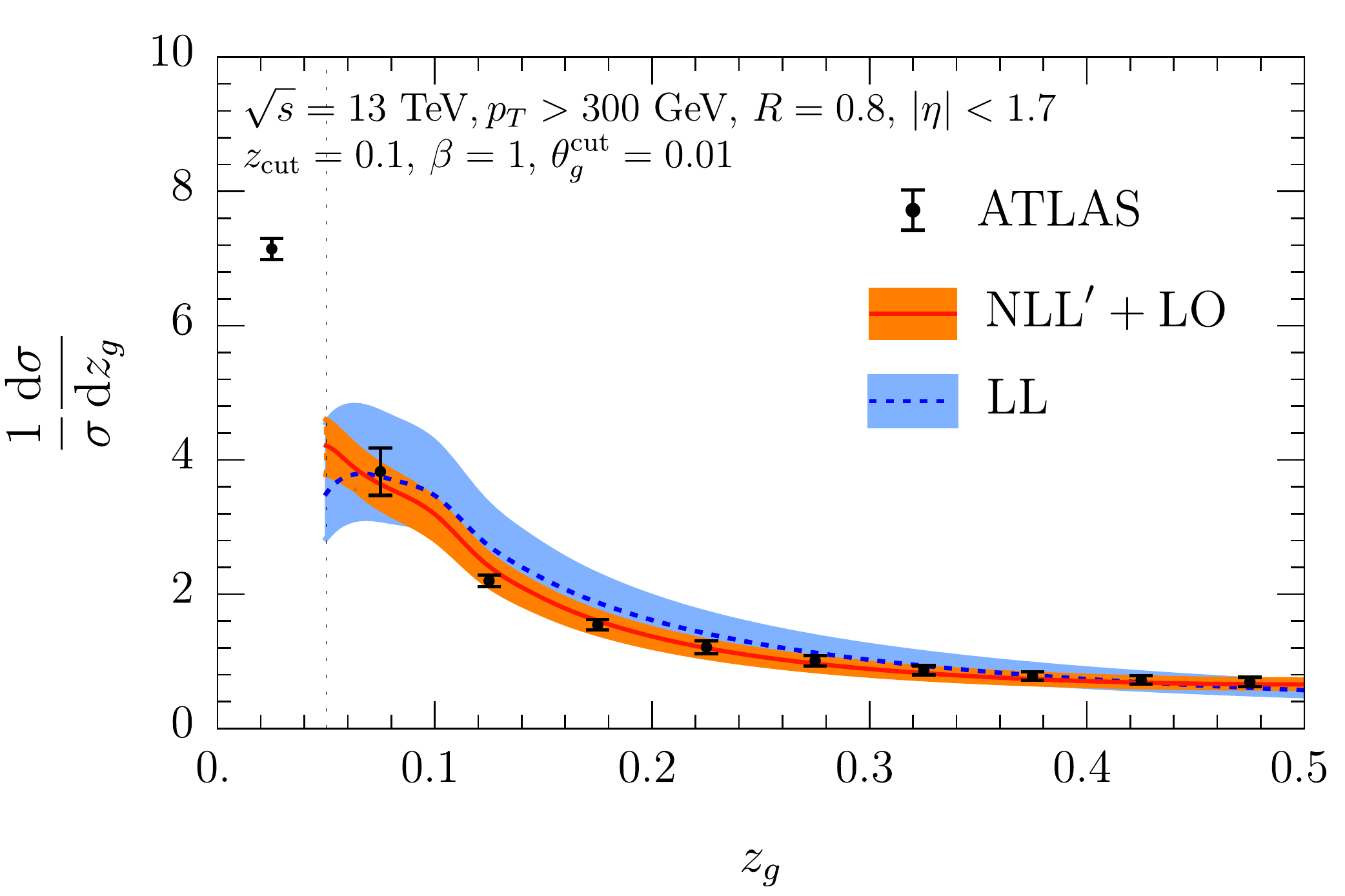}
  \caption{Comparison to ATLAS results~\cite{Aad:2019vyi} for $\beta=1$.~\label{fig:data_beta1}}
\end{figure}
\begin{figure}[b]
  \includegraphics[width = 0.48 \textwidth]{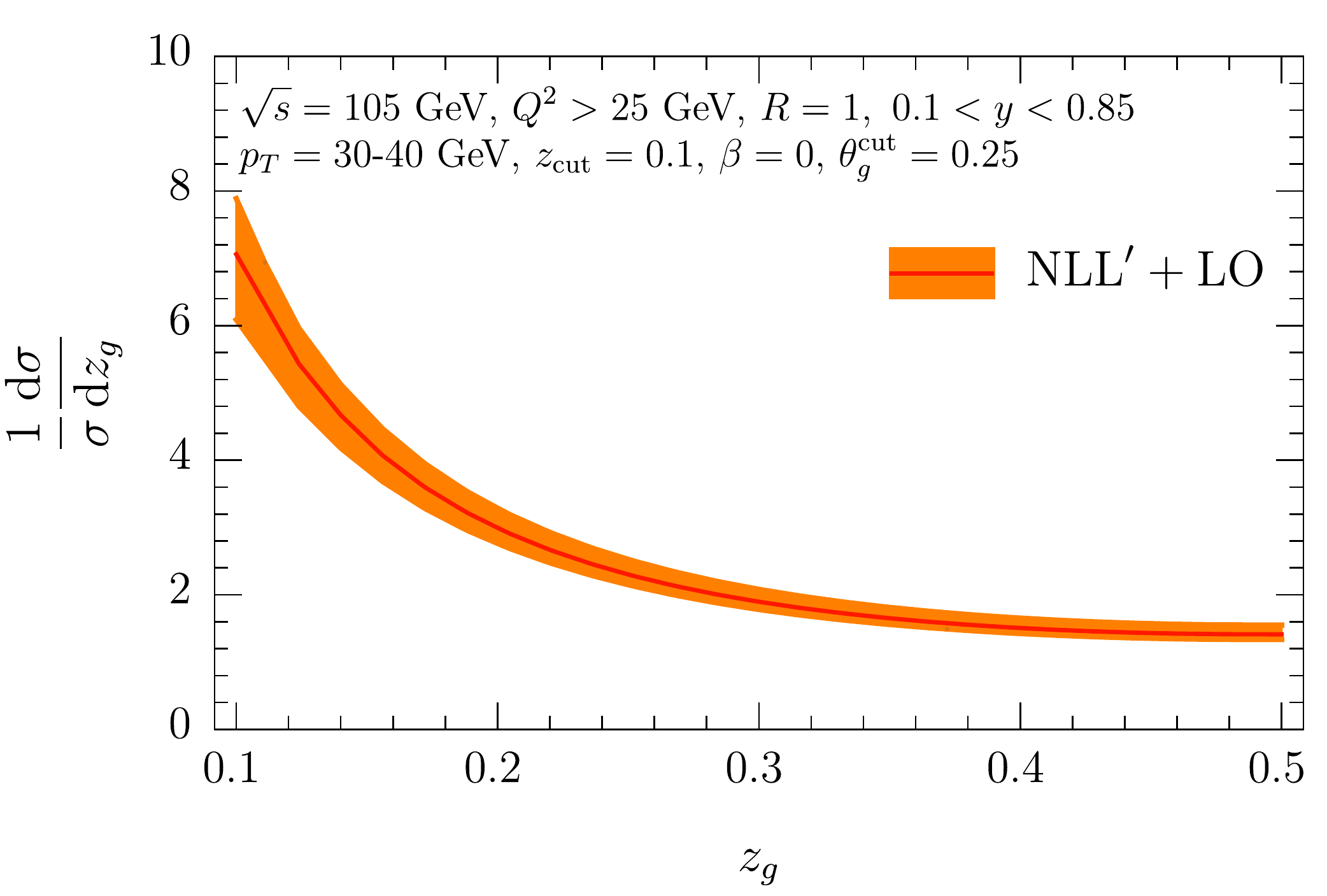}
  \caption{Prediction for the Electron-Ion Collider.~\label{fig:EIC}}
\end{figure}

Next, we compare to the experimental results from ATLAS~\cite{Aad:2019vyi}, ALICE~\cite{Mulligan:2020cnp} and STAR~\cite{Adam:2020kug} for $\beta=0$ in Fig.~\ref{fig:data_beta0}. We normalize our results to the data~\footnote{The ALICE normalization is not normalized by a few percent, because of their treatment of jets that never pass the soft drop condition.} and impose the same cut on $\theta_g$. The hadronization effects in Pythia for ALICE and STAR kinematics (not shown) are much more sizable than in Fig.~\ref{fig:Pythia}, in accord with the larger perturbative uncertainties. Nevertheless, we find very good agreement even for these relatively low jet transverse momenta. We note that the CMS result of Ref.~\cite{Sirunyan:2017bsd} is not unfolded, prohibiting a direct comparison. As a representative example, we show the LL QCD scale uncertainty band in the left panel of Fig.~\ref{fig:data_beta0} which is significantly larger than at NLL$'$. This implies that the NLL$'$ accuracy achieved in this work is needed to match the current experimental precision.

Next, we compare in Fig.~\ref{fig:data_beta1} our results to ATLAS measurement~\cite{Aad:2019vyi} for $\beta=1$, as an example. We normalize our results to the data in the region to the right of the dotted line,  as our prediction for the left most data point are very sensitive to nonperturbative effects. Note that this was not needed for $\beta=0$, where $z_{\rm cut}$ provides a lower bound on $z_g$. We observe excellent agreement! In this case our NLL$'$+LO prediction is substantially better than the LL result.

Lastly, we present predictions for jet kinematics at the future EIC in Fig.~\ref{fig:EIC}. We consider jets reconstructed in the laboratory frame $ep\to e+{\rm jet}+X$ for typical EIC kinematics~\cite{Arratia:2019vju,Page:2019gbf} with cuts on the photon virtuality $Q^2$ and the inelasticity $y$ as indicated in the figure. The clean environment at the EIC will allow studies of hadronization effects, and $z_g$ measurements in single- and di-jet events can help improve our understanding of quark/gluon differences (see also Fig.~\ref{fig:spin_color}).

{\it Conclusions.} In this work we have presented a calculation of the soft drop groomed momentum sharing fraction $z_g$ at NLL$'$+LO accuracy. This Sudakov-safe jet substructure observable, which probes the hard branching process inside the jet, constitutes the most direct measurement of the QCD splitting function. Our framework allows for a systematic extension beyond the previously achieved LL accuracy, yielding qualitatively different results for $\beta<0$ than in an earlier study, and provides the first meaningful assessment of theoretical uncertainties. We also show that $z_g$ probes the QCD splitting function beyond the leading $1/z_g$ dependence, indicating sensitivity to the spin of the particle that initiates the jet. Our calculations indicate that this effect is sufficiently large to be probed experimentally. The momentum sharing fraction $z_g$ is one of the hallmark observables in the field of jet substructure and has been measured by several experimental collaborations at the LHC and RHIC. We compared to the available experimental data and found very good agreement with our purely perturbative calculation. In addition, we provided predictions for the future Electron-Ion Collider. Our precise calculations reduced uncertainties and changed the shape of the prediction, bringing it in excellent agreement with the data.

{\it Acknowledgements.} We thank Yi Chen, Raghav Elayavalli, Matt LeBlanc, Yen-Jie Lee, Ezra Lesser, James Mulligan, Ben Nachman, Mateusz Ploskon, Jennifer Roloff, and Marta Verweij for helpful discussions. We also thank Andrew Larkoski and Jesse Thaler for feedback on our manuscript. PC and WW are supported by the ERC grant ERC-STG-2015-677323, the NWO projectruimte 680-91-122 and the D-ITP consortium, a program of NWO that is funded by the Dutch Ministry of Education, Culture and Science (OCW). KL is supported by the US Department of Energy, Office of Nuclear Physics. FR is supported by the US Department of Energy under Contract No. DE-AC02-05CH11231 and the LDRD Program of LBNL.

\bibliographystyle{utphys}
\bibliography{main.bib}

\providecommand{\href}[2]{#2}\begingroup\raggedright\begin{thebibliography}{10}

\bibitem{Larkoski:2017jix}
A.~J. Larkoski, I.~Moult, and B.~Nachman, ``{Jet Substructure at the Large
  Hadron Collider: A Review of Recent Advances in Theory and Machine
  Learning},'' \href{http://dx.doi.org/10.1016/j.physrep.2019.11.001}{{\em
  Phys. Rept.} {\bfseries 841} (2020) 1--63},
  \href{http://arxiv.org/abs/1709.04464}{{\ttfamily arXiv:1709.04464
  [hep-ph]}}.

\bibitem{Asquith:2018igt}
R.~Kogler {\em et~al.}, ``{Jet Substructure at the Large Hadron Collider:
  Experimental Review},''
  \href{http://dx.doi.org/10.1103/RevModPhys.91.045003}{{\em Rev. Mod. Phys.}
  {\bfseries 91} no.~4, (2019) 045003},
  \href{http://arxiv.org/abs/1803.06991}{{\ttfamily arXiv:1803.06991
  [hep-ex]}}.

\bibitem{Marzani:2019hun}
S.~Marzani, G.~Soyez, and M.~Spannowsky,
  \href{http://dx.doi.org/10.1007/978-3-030-15709-8}{{\em {Looking inside jets:
  an introduction to jet substructure and boosted-object phenomenology}}},
  vol.~958.
\newblock Springer, 2019.
\newblock \href{http://arxiv.org/abs/1901.10342}{{\ttfamily arXiv:1901.10342
  [hep-ph]}}.

\bibitem{Krohn:2009th}
D.~Krohn, J.~Thaler, and L.-T. Wang, ``{Jet Trimming},''
  \href{http://dx.doi.org/10.1007/JHEP02(2010)084}{{\em JHEP} {\bfseries 02}
  (2010) 084}, \href{http://arxiv.org/abs/0912.1342}{{\ttfamily arXiv:0912.1342
  [hep-ph]}}.

\bibitem{Ellis:2009me}
S.~D. Ellis, C.~K. Vermilion, and J.~R. Walsh, ``{Recombination Algorithms and
  Jet Substructure: Pruning as a Tool for Heavy Particle Searches},''
  \href{http://dx.doi.org/10.1103/PhysRevD.81.094023}{{\em Phys. Rev. D}
  {\bfseries 81} (2010) 094023},
  \href{http://arxiv.org/abs/0912.0033}{{\ttfamily arXiv:0912.0033 [hep-ph]}}.

\bibitem{Dasgupta:2013ihk}
M.~Dasgupta, A.~Fregoso, S.~Marzani, and G.~P. Salam, ``{Towards an
  understanding of jet substructure},''
  \href{http://dx.doi.org/10.1007/JHEP09(2013)029}{{\em JHEP} {\bfseries 09}
  (2013) 029},
\href{http://arxiv.org/abs/1307.0007}{{\ttfamily arXiv:1307.0007 [hep-ph]}}.

\bibitem{Cacciari:2014gra}
M.~Cacciari, G.~P. Salam, and G.~Soyez, ``{SoftKiller, a particle-level pileup
  removal method},''
  \href{http://dx.doi.org/10.1140/epjc/s10052-015-3267-2}{{\em Eur. Phys. J. C}
  {\bfseries 75} no.~2, (2015) 59},
  \href{http://arxiv.org/abs/1407.0408}{{\ttfamily arXiv:1407.0408 [hep-ph]}}.

\bibitem{Larkoski:2014wba}
A.~J. Larkoski, S.~Marzani, G.~Soyez, and J.~Thaler, ``{Soft Drop},''
  \href{http://dx.doi.org/10.1007/JHEP05(2014)146}{{\em JHEP} {\bfseries 05}
  (2014) 146},
\href{http://arxiv.org/abs/1402.2657}{{\ttfamily arXiv:1402.2657 [hep-ph]}}.

\bibitem{Aaboud:2017qwh}
{\bfseries ATLAS} Collaboration, M.~Aaboud {\em et~al.}, ``{Measurement of the
  Soft-Drop Jet Mass in pp Collisions at $\sqrt{s} = 13$ TeV with the ATLAS
  Detector},'' \href{http://dx.doi.org/10.1103/PhysRevLett.121.092001}{{\em
  Phys. Rev. Lett.} {\bfseries 121} no.~9, (2018) 092001},
\href{http://arxiv.org/abs/1711.08341}{{\ttfamily arXiv:1711.08341 [hep-ex]}}.

\bibitem{Sirunyan:2018xdh}
{\bfseries CMS} Collaboration, A.~M. Sirunyan {\em et~al.}, ``{Measurements of
  the differential jet cross section as a function of the jet mass in dijet
  events from proton-proton collisions at $ \sqrt{s}=13 $ TeV},''
  \href{http://dx.doi.org/10.1007/JHEP11(2018)113}{{\em JHEP} {\bfseries 11}
  (2018) 113},
\href{http://arxiv.org/abs/1807.05974}{{\ttfamily arXiv:1807.05974 [hep-ex]}}.

\bibitem{Aad:2019vyi}
{\bfseries ATLAS} Collaboration, G.~Aad {\em et~al.}, ``{Measurement of
  soft-drop jet observables in $pp$ collisions with the ATLAS detector at
  $\sqrt {s}$ =13 TeV},''
  \href{http://dx.doi.org/10.1103/PhysRevD.101.052007}{{\em Phys. Rev. D}
  {\bfseries 101} no.~5, (2020) 052007},
  \href{http://arxiv.org/abs/1912.09837}{{\ttfamily arXiv:1912.09837
  [hep-ex]}}.

\bibitem{Dokshitzer:1997in}
Y.~L. Dokshitzer, G.~D. Leder, S.~Moretti, and B.~R. Webber, ``{Better jet
  clustering algorithms},''
  \href{http://dx.doi.org/10.1088/1126-6708/1997/08/001}{{\em JHEP} {\bfseries
  08} (1997) 001},
\href{http://arxiv.org/abs/hep-ph/9707323}{{\ttfamily arXiv:hep-ph/9707323
  [hep-ph]}}.

\bibitem{Wobisch:1998wt}
M.~Wobisch and T.~Wengler, ``{Hadronization corrections to jet cross-sections
  in deep inelastic scattering},'' in {\em {Monte Carlo generators for HERA
  physics. Proceedings, Workshop, Hamburg, Germany, 1998-1999}}, pp.~270--279.
\newblock 1998.
\newblock \href{http://arxiv.org/abs/hep-ph/9907280}{{\ttfamily
  arXiv:hep-ph/9907280 [hep-ph]}}.
\newblock
\url{http://inspirehep.net/record/484872/files/arXiv:hep-ph_9907280.pdf}.
\newblock

\bibitem{Altarelli:1977zs}
G.~Altarelli and G.~Parisi, ``{Asymptotic Freedom in Parton Language},''
\href{http://dx.doi.org/10.1016/0550-3213(77)90384-4}{{\em Nucl. Phys.}
  {\bfseries B126} (1977) 298--318}.

\bibitem{Acharya:2019djg}
{\bfseries ALICE} Collaboration, S.~Acharya {\em et~al.}, ``{Exploration of jet
  substructure using iterative declustering in pp and Pb-Pb collisions at LHC
  energies},''
\href{http://arxiv.org/abs/1905.02512}{{\ttfamily arXiv:1905.02512 [nucl-ex]}}.

\bibitem{Mulligan:2020cnp}
{\bfseries ALICE} Collaboration, J.~Mulligan, ``{Jet substructure measurements
  in $\mathrm{p\kern-0.05em p}$ and $\mbox{Pb-Pb}$ collisions at
  $\sqrt{s_{\mathrm{NN}}}=5.02$ TeV with ALICE},''
  \href{http://arxiv.org/abs/2009.07172}{{\ttfamily arXiv:2009.07172
  [nucl-ex]}}. \url{https://alice-figure.web.cern.ch/node/16086}.

\bibitem{Sirunyan:2017bsd}
{\bfseries CMS} Collaboration, A.~M. Sirunyan {\em et~al.}, ``{Measurement of
  the Splitting Function in $pp$ and Pb-Pb Collisions at
  $\sqrt{s_{_{\mathrm{NN}}}} =$ 5.02 TeV},''
  \href{http://dx.doi.org/10.1103/PhysRevLett.120.142302}{{\em Phys. Rev.
  Lett.} {\bfseries 120} no.~14, (2018) 142302},
\href{http://arxiv.org/abs/1708.09429}{{\ttfamily arXiv:1708.09429 [nucl-ex]}}.

\bibitem{Sirunyan:2018asm}
{\bfseries CMS} Collaboration, A.~M. Sirunyan {\em et~al.}, ``{Measurement of
  jet substructure observables in $\mathrm{t\overline{t}}$ events from
  proton-proton collisions at $\sqrt{s}=$ 13TeV},''
  \href{http://dx.doi.org/10.1103/PhysRevD.98.092014}{{\em Phys. Rev. D}
  {\bfseries 98} no.~9, (2018) 092014},
  \href{http://arxiv.org/abs/1808.07340}{{\ttfamily arXiv:1808.07340
  [hep-ex]}}.

\bibitem{Adam:2020kug}
{\bfseries STAR} Collaboration, J.~Adam {\em et~al.}, ``{Measurement of groomed
  jet substructure observables in p+p collisions at $\sqrt {s}$ =200 GeV with
  STAR},'' \href{http://dx.doi.org/10.1016/j.physletb.2020.135846}{{\em Phys.
  Lett. B} {\bfseries 811} (2020) 135846},
  \href{http://arxiv.org/abs/2003.02114}{{\ttfamily arXiv:2003.02114
  [hep-ex]}}.

\bibitem{Larkoski:2017bvj}
A.~Larkoski, S.~Marzani, J.~Thaler, A.~Tripathee, and W.~Xue, ``{Exposing the
  QCD Splitting Function with CMS Open Data},''
  \href{http://dx.doi.org/10.1103/PhysRevLett.119.132003}{{\em Phys. Rev.
  Lett.} {\bfseries 119} no.~13, (2017) 132003},
  \href{http://arxiv.org/abs/1704.05066}{{\ttfamily arXiv:1704.05066
  [hep-ph]}}.

\bibitem{Tripathee:2017ybi}
A.~Tripathee, W.~Xue, A.~Larkoski, S.~Marzani, and J.~Thaler, ``{Jet
  Substructure Studies with CMS Open Data},''
  \href{http://dx.doi.org/10.1103/PhysRevD.96.074003}{{\em Phys. Rev. D}
  {\bfseries 96} no.~7, (2017) 074003},
  \href{http://arxiv.org/abs/1704.05842}{{\ttfamily arXiv:1704.05842
  [hep-ph]}}.

\bibitem{Mehtar-Tani:2016aco}
Y.~Mehtar-Tani and K.~Tywoniuk, ``{Groomed jets in heavy-ion collisions:
  sensitivity to medium-induced bremsstrahlung},''
  \href{http://dx.doi.org/10.1007/JHEP04(2017)125}{{\em JHEP} {\bfseries 04}
  (2017) 125}, \href{http://arxiv.org/abs/1610.08930}{{\ttfamily
  arXiv:1610.08930 [hep-ph]}}.

\bibitem{Chien:2016led}
Y.-T. Chien and I.~Vitev, ``{Probing the Hardest Branching within Jets in
  Heavy-Ion Collisions},''
  \href{http://dx.doi.org/10.1103/PhysRevLett.119.112301}{{\em Phys. Rev.
  Lett.} {\bfseries 119} no.~11, (2017) 112301},
  \href{http://arxiv.org/abs/1608.07283}{{\ttfamily arXiv:1608.07283
  [hep-ph]}}.

\bibitem{Chang:2017gkt}
N.-B. Chang, S.~Cao, and G.-Y. Qin, ``{Probing medium-induced jet splitting and
  energy loss in heavy-ion collisions},''
  \href{http://dx.doi.org/10.1016/j.physletb.2018.04.019}{{\em Phys. Lett. B}
  {\bfseries 781} (2018) 423--432},
  \href{http://arxiv.org/abs/1707.03767}{{\ttfamily arXiv:1707.03767
  [hep-ph]}}.

\bibitem{Li:2017wwc}
H.~T. Li and I.~Vitev, ``{Inverting the mass hierarchy of jet quenching effects
  with prompt $b$-jet substructure},''
  \href{http://dx.doi.org/10.1016/j.physletb.2019.04.052}{{\em Phys. Lett. B}
  {\bfseries 793} (2019) 259--264},
  \href{http://arxiv.org/abs/1801.00008}{{\ttfamily arXiv:1801.00008
  [hep-ph]}}.

\bibitem{Milhano:2017nzm}
G.~Milhano, U.~A. Wiedemann, and K.~C. Zapp, ``{Sensitivity of jet substructure
  to jet-induced medium response},''
  \href{http://dx.doi.org/10.1016/j.physletb.2018.01.029}{{\em Phys. Lett. B}
  {\bfseries 779} (2018) 409--413},
  \href{http://arxiv.org/abs/1707.04142}{{\ttfamily arXiv:1707.04142
  [hep-ph]}}.

\bibitem{KunnawalkamElayavalli:2017hxo}
R.~Kunnawalkam~Elayavalli and K.~C. Zapp, ``{Medium response in JEWEL and its
  impact on jet shape observables in heavy ion collisions},''
  \href{http://dx.doi.org/10.1007/JHEP07(2017)141}{{\em JHEP} {\bfseries 07}
  (2017) 141}, \href{http://arxiv.org/abs/1707.01539}{{\ttfamily
  arXiv:1707.01539 [hep-ph]}}.

\bibitem{Casalderrey-Solana:2019ubu}
J.~Casalderrey-Solana, G.~Milhano, D.~Pablos, and K.~Rajagopal, ``{Modification
  of Jet Substructure in Heavy Ion Collisions as a Probe of the Resolution
  Length of Quark-Gluon Plasma},''
  \href{http://dx.doi.org/10.1007/JHEP01(2020)044}{{\em JHEP} {\bfseries 01}
  (2020) 044}, \href{http://arxiv.org/abs/1907.11248}{{\ttfamily
  arXiv:1907.11248 [hep-ph]}}.

\bibitem{Caucal:2019uvr}
P.~Caucal, E.~Iancu, and G.~Soyez, ``{Deciphering the $z_g$ distribution in
  ultrarelativistic heavy ion collisions},''
  \href{http://dx.doi.org/10.1007/JHEP10(2019)273}{{\em JHEP} {\bfseries 10}
  (2019) 273}, \href{http://arxiv.org/abs/1907.04866}{{\ttfamily
  arXiv:1907.04866 [hep-ph]}}.

\bibitem{AbdulKhalek:2021gbh}
R.~Abdul~Khalek {\em et~al.}, ``{Science Requirements and Detector Concepts for
  the Electron-Ion Collider: EIC Yellow Report},''
  \href{http://arxiv.org/abs/2103.05419}{{\ttfamily arXiv:2103.05419
  [physics.ins-det]}}.

\bibitem{Larkoski:2015lea}
A.~J. Larkoski, S.~Marzani, and J.~Thaler, ``{Sudakov Safety in Perturbative
  QCD},'' \href{http://dx.doi.org/10.1103/PhysRevD.91.111501}{{\em Phys. Rev.
  D} {\bfseries 91} no.~11, (2015) 111501},
  \href{http://arxiv.org/abs/1502.01719}{{\ttfamily arXiv:1502.01719
  [hep-ph]}}.

\bibitem{Larkoski:2013paa}
A.~J. Larkoski and J.~Thaler, ``{Unsafe but Calculable: Ratios of Angularities
  in Perturbative QCD},'' \href{http://dx.doi.org/10.1007/JHEP09(2013)137}{{\em
  JHEP} {\bfseries 09} (2013) 137},
  \href{http://arxiv.org/abs/1307.1699}{{\ttfamily arXiv:1307.1699 [hep-ph]}}.

\bibitem{Bauer:2000ew}
C.~W. Bauer, S.~Fleming, and M.~E. Luke, ``{Summing Sudakov logarithms in $B
  \to X_s \gamma$ in effective field theory},''
  \href{http://dx.doi.org/10.1103/PhysRevD.63.014006}{{\em Phys. Rev.}
  {\bfseries D63} (2000) 014006},
\href{http://arxiv.org/abs/hep-ph/0005275}{{\ttfamily arXiv:hep-ph/0005275
  [hep-ph]}}.

\bibitem{Bauer:2000yr}
C.~W. Bauer, S.~Fleming, D.~Pirjol, and I.~W. Stewart, ``{An Effective field
  theory for collinear and soft gluons: Heavy to light decays},''
  \href{http://dx.doi.org/10.1103/PhysRevD.63.114020}{{\em Phys. Rev.}
  {\bfseries D63} (2001) 114020},
\href{http://arxiv.org/abs/hep-ph/0011336}{{\ttfamily arXiv:hep-ph/0011336
  [hep-ph]}}.

\bibitem{Bauer:2001yt}
C.~W. Bauer, D.~Pirjol, and I.~W. Stewart, ``{Soft collinear factorization in
  effective field theory},''
  \href{http://dx.doi.org/10.1103/PhysRevD.65.054022}{{\em Phys. Rev.}
  {\bfseries D65} (2002) 054022},
\href{http://arxiv.org/abs/hep-ph/0109045}{{\ttfamily arXiv:hep-ph/0109045
  [hep-ph]}}.

\bibitem{Bauer:2002nz}
C.~W. Bauer, S.~Fleming, D.~Pirjol, I.~Z. Rothstein, and I.~W. Stewart, ``{Hard
  scattering factorization from effective field theory},''
  \href{http://dx.doi.org/10.1103/PhysRevD.66.014017}{{\em Phys. Rev.}
  {\bfseries D66} (2002) 014017},
\href{http://arxiv.org/abs/hep-ph/0202088}{{\ttfamily arXiv:hep-ph/0202088
  [hep-ph]}}.

\bibitem{Beneke:2002ph}
M.~Beneke, A.~P. Chapovsky, M.~Diehl, and T.~Feldmann, ``{Soft collinear
  effective theory and heavy to light currents beyond leading power},''
  \href{http://dx.doi.org/10.1016/S0550-3213(02)00687-9}{{\em Nucl. Phys.}
  {\bfseries B643} (2002) 431--476},
\href{http://arxiv.org/abs/hep-ph/0206152}{{\ttfamily arXiv:hep-ph/0206152
  [hep-ph]}}.

\bibitem{Dasgupta:2001sh}
M.~Dasgupta and G.~P. Salam, ``{Resummation of nonglobal QCD observables},''
  \href{http://dx.doi.org/10.1016/S0370-2693(01)00725-0}{{\em Phys. Lett.}
  {\bfseries B512} (2001) 323--330},
\href{http://arxiv.org/abs/hep-ph/0104277}{{\ttfamily arXiv:hep-ph/0104277
  [hep-ph]}}.

\bibitem{Hoang:2019ceu}
A.~H. Hoang, S.~Mantry, A.~Pathak, and I.~W. Stewart, ``{Nonperturbative
  Corrections to Soft Drop Jet Mass},''
\href{http://arxiv.org/abs/1906.11843}{{\ttfamily arXiv:1906.11843 [hep-ph]}}.

\bibitem{Pathak:2020iue}
A.~Pathak, I.~W. Stewart, V.~Vaidya, and L.~Zoppi, ``{EFT for Soft Drop Double
  Differential Cross Section},''
  \href{http://arxiv.org/abs/2012.15568}{{\ttfamily arXiv:2012.15568
  [hep-ph]}}.

\bibitem{Frye:2016aiz}
C.~Frye, A.~J. Larkoski, M.~D. Schwartz, and K.~Yan, ``{Factorization for
  groomed jet substructure beyond the next-to-leading logarithm},''
  \href{http://dx.doi.org/10.1007/JHEP07(2016)064}{{\em JHEP} {\bfseries 07}
  (2016) 064},
\href{http://arxiv.org/abs/1603.09338}{{\ttfamily arXiv:1603.09338 [hep-ph]}}.

\bibitem{Marzani:2017mva}
S.~Marzani, L.~Schunk, and G.~Soyez, ``{A study of jet mass distributions with
  grooming},'' \href{http://dx.doi.org/10.1007/JHEP07(2017)132}{{\em JHEP}
  {\bfseries 07} (2017) 132},
\href{http://arxiv.org/abs/1704.02210}{{\ttfamily arXiv:1704.02210 [hep-ph]}}.

\bibitem{Larkoski:2017cqq}
A.~J. Larkoski, I.~Moult, and D.~Neill, ``{Factorization and Resummation for
  Groomed Multi-Prong Jet Shapes},''
  \href{http://dx.doi.org/10.1007/JHEP02(2018)144}{{\em JHEP} {\bfseries 02}
  (2018) 144},
\href{http://arxiv.org/abs/1710.00014}{{\ttfamily arXiv:1710.00014 [hep-ph]}}.

\bibitem{Kang:2018jwa}
Z.-B. Kang, K.~Lee, X.~Liu, and F.~Ringer, ``{The groomed and ungroomed jet
  mass distribution for inclusive jet production at the LHC},''
  \href{http://dx.doi.org/10.1007/JHEP10(2018)137}{{\em JHEP} {\bfseries 10}
  (2018) 137},
\href{http://arxiv.org/abs/1803.03645}{{\ttfamily arXiv:1803.03645 [hep-ph]}}.

\bibitem{Kardos:2018kth}
A.~Kardos, G.~Somogyi, and Z.~Tr{\'o}cs{\'a}nyi, ``{Soft-drop event shapes in
  electron--positron annihilation at next-to-next-to-leading order accuracy},''
  \href{http://dx.doi.org/10.1016/j.physletb.2018.10.014}{{\em Phys. Lett.}
  {\bfseries B786} (2018) 313--318},
\href{http://arxiv.org/abs/1807.11472}{{\ttfamily arXiv:1807.11472 [hep-ph]}}.

\bibitem{Lee:2019lge}
C.~Lee, P.~Shrivastava, and V.~Vaidya, ``{Predictions for energy correlators
  probing substructure of groomed heavy quark jets},''
  \href{http://dx.doi.org/10.1007/JHEP09(2019)045}{{\em JHEP} {\bfseries 09}
  (2019) 045},
\href{http://arxiv.org/abs/1901.09095}{{\ttfamily arXiv:1901.09095 [hep-ph]}}.

\bibitem{Chien:2019osu}
Y.-T. Chien and I.~W. Stewart, ``{Collinear Drop},''
  \href{http://dx.doi.org/10.1007/JHEP06(2020)064}{{\em JHEP} {\bfseries 06}
  (2020) 064}, \href{http://arxiv.org/abs/1907.11107}{{\ttfamily
  arXiv:1907.11107 [hep-ph]}}.

\bibitem{Larkoski:2020wgx}
A.~J. Larkoski, ``{Improving the understanding of jet grooming in perturbation
  theory},'' \href{http://dx.doi.org/10.1007/JHEP09(2020)072}{{\em JHEP}
  {\bfseries 09} (2020) 072}, \href{http://arxiv.org/abs/2006.14680}{{\ttfamily
  arXiv:2006.14680 [hep-ph]}}.

\bibitem{Kardos:2020gty}
A.~Kardos, A.~J. Larkoski, and Z.~Tr\'ocs\'anyi, ``{Groomed jet mass at high
  precision},'' \href{http://dx.doi.org/10.1016/j.physletb.2020.135704}{{\em
  Phys. Lett. B} {\bfseries 809} (2020) 135704},
  \href{http://arxiv.org/abs/2002.00942}{{\ttfamily arXiv:2002.00942
  [hep-ph]}}.

\bibitem{Anderle:2020mxj}
D.~Anderle, M.~Dasgupta, B.~K. El-Menoufi, J.~Helliwell, and M.~Guzzi,
  ``{Groomed jet mass as a direct probe of collinear parton dynamics},''
  \href{http://dx.doi.org/10.1140/epjc/s10052-020-8411-y}{{\em Eur. Phys. J. C}
  {\bfseries 80} no.~9, (2020) 827},
  \href{http://arxiv.org/abs/2007.10355}{{\ttfamily arXiv:2007.10355
  [hep-ph]}}.

\bibitem{Mehtar-Tani:2020oux}
Y.~Mehtar-Tani, A.~Soto-Ontoso, and K.~Tywoniuk, ``{Tagging boosted hadronic
  objects with dynamical grooming},''
  \href{http://dx.doi.org/10.1103/PhysRevD.102.114013}{{\em Phys. Rev. D}
  {\bfseries 102} (2020) 114013},
  \href{http://arxiv.org/abs/2005.07584}{{\ttfamily arXiv:2005.07584
  [hep-ph]}}.

\bibitem{Baron:2020xoi}
J.~Baron, D.~Reichelt, S.~Schumann, N.~Schwanemann, and V.~Theeuwes,
  ``{Soft-drop grooming for hadronic event shapes},''
  \href{http://arxiv.org/abs/2012.09574}{{\ttfamily arXiv:2012.09574
  [hep-ph]}}.

\bibitem{Makris:2021drz}
Y.~Makris, ``{Revisiting the role of grooming in DIS},''
  \href{http://arxiv.org/abs/2101.02708}{{\ttfamily arXiv:2101.02708
  [hep-ph]}}.

\bibitem{Caucal:2021bae}
P.~Caucal, A.~Soto-Ontoso, and A.~Takacs, ``{Dynamical grooming meets LHC
  data},'' \href{http://arxiv.org/abs/2103.06566}{{\ttfamily arXiv:2103.06566
  [hep-ph]}}.

\bibitem{Cal:2019gxa}
P.~Cal, D.~Neill, F.~Ringer, and W.~J. Waalewijn, ``{Calculating the angle
  between jet axes},'' \href{http://dx.doi.org/10.1007/JHEP04(2020)211}{{\em
  JHEP} {\bfseries 04} (2020) 211},
  \href{http://arxiv.org/abs/1911.06840}{{\ttfamily arXiv:1911.06840
  [hep-ph]}}.

\bibitem{Cal:2020flh}
P.~Cal, K.~Lee, F.~Ringer, and W.~J. Waalewijn, ``{Jet energy drop},''
  \href{http://dx.doi.org/10.1007/JHEP11(2020)012}{{\em JHEP} {\bfseries 11}
  (2020) 012}, \href{http://arxiv.org/abs/2007.12187}{{\ttfamily
  arXiv:2007.12187 [hep-ph]}}.

\bibitem{Caletti:2021oor}
S.~Caletti, O.~Fedkevych, S.~Marzani, D.~Reichelt, S.~Schumann, G.~Soyez, and
  V.~Theeuwes, ``{Jet Angularities in Z+jet production at the LHC},''
  \href{http://arxiv.org/abs/2104.06920}{{\ttfamily arXiv:2104.06920
  [hep-ph]}}.

\bibitem{Aversa:1988vb}
F.~Aversa, P.~Chiappetta, M.~Greco, and J.~P. Guillet, ``{QCD Corrections to
  Parton-Parton Scattering Processes},''
\href{http://dx.doi.org/10.1016/0550-3213(89)90288-5}{{\em Nucl. Phys.}
  {\bfseries B327} (1989) 105}.

\bibitem{Jager:2002xm}
B.~Jager, A.~Schafer, M.~Stratmann, and W.~Vogelsang, ``{Next-to-leading order
  QCD corrections to high $p_T$ pion production in longitudinally polarized pp
  collisions},'' \href{http://dx.doi.org/10.1103/PhysRevD.67.054005}{{\em Phys.
  Rev.} {\bfseries D67} (2003) 054005},
\href{http://arxiv.org/abs/hep-ph/0211007}{{\ttfamily arXiv:hep-ph/0211007
  [hep-ph]}}.

\bibitem{Mukherjee:2012uz}
A.~Mukherjee and W.~Vogelsang, ``{Jet production in (un)polarized pp
  collisions: dependence on jet algorithm},''
  \href{http://dx.doi.org/10.1103/PhysRevD.86.094009}{{\em Phys. Rev.}
  {\bfseries D86} (2012) 094009},
\href{http://arxiv.org/abs/1209.1785}{{\ttfamily arXiv:1209.1785 [hep-ph]}}.

\bibitem{Dasgupta:2014yra}
M.~Dasgupta, F.~Dreyer, G.~P. Salam, and G.~Soyez, ``{Small-radius jets to all
  orders in QCD},'' \href{http://dx.doi.org/10.1007/JHEP04(2015)039}{{\em JHEP}
  {\bfseries 04} (2015) 039},
\href{http://arxiv.org/abs/1411.5182}{{\ttfamily arXiv:1411.5182 [hep-ph]}}.

\bibitem{Kaufmann:2015hma}
T.~Kaufmann, A.~Mukherjee, and W.~Vogelsang, ``{Hadron Fragmentation Inside
  Jets in Hadronic Collisions},''
  \href{http://dx.doi.org/10.1103/PhysRevD.92.054015}{{\em Phys. Rev.}
  {\bfseries D92} (2015) 054015},
\href{http://arxiv.org/abs/1506.01415}{{\ttfamily arXiv:1506.01415 [hep-ph]}}.

\bibitem{Kang:2016mcy}
Z.-B. Kang, F.~Ringer, and I.~Vitev, ``{The semi-inclusive jet function in SCET
  and small radius resummation for inclusive jet production},''
  \href{http://dx.doi.org/10.1007/JHEP10(2016)125}{{\em JHEP} {\bfseries 10}
  (2016) 125},
\href{http://arxiv.org/abs/1606.06732}{{\ttfamily arXiv:1606.06732 [hep-ph]}}.

\bibitem{Dai:2016hzf}
L.~Dai, C.~Kim, and A.~K. Leibovich, ``{Fragmentation of a Jet with Small
  Radius},'' \href{http://dx.doi.org/10.1103/PhysRevD.94.114023}{{\em Phys.
  Rev.} {\bfseries D94} no.~11, (2016) 114023},
\href{http://arxiv.org/abs/1606.07411}{{\ttfamily arXiv:1606.07411 [hep-ph]}}.

\bibitem{Sjostrand:2014zea}
T.~Sj{\"o}strand, S.~Ask, J.~R. Christiansen, R.~Corke, N.~Desai, P.~Ilten,
  S.~Mrenna, S.~Prestel, C.~O. Rasmussen, and P.~Z. Skands, ``{An Introduction
  to PYTHIA 8.2},'' \href{http://dx.doi.org/10.1016/j.cpc.2015.01.024}{{\em
  Comput. Phys. Commun.} {\bfseries 191} (2015) 159--177},
\href{http://arxiv.org/abs/1410.3012}{{\ttfamily arXiv:1410.3012 [hep-ph]}}.

\bibitem{Kang:2019prh}
Z.-B. Kang, K.~Lee, X.~Liu, D.~Neill, and F.~Ringer, ``{The soft drop groomed
  jet radius at NLL},'' \href{http://dx.doi.org/10.1007/JHEP02(2020)054}{{\em
  JHEP} {\bfseries 02} (2020) 054},
  \href{http://arxiv.org/abs/1908.01783}{{\ttfamily arXiv:1908.01783
  [hep-ph]}}.

\bibitem{Banfi:2002hw}
A.~Banfi, G.~Marchesini, and G.~Smye, ``{Away from jet energy flow},''
  \href{http://dx.doi.org/10.1088/1126-6708/2002/08/006}{{\em JHEP} {\bfseries
  08} (2002) 006}, \href{http://arxiv.org/abs/hep-ph/0206076}{{\ttfamily
  arXiv:hep-ph/0206076}}.

\bibitem{Hornig:2011iu}
A.~Hornig, C.~Lee, I.~W. Stewart, J.~R. Walsh, and S.~Zuberi, ``{Non-global
  Structure of the $O({\alpha}_s^2)$ Dijet Soft Function},''
  \href{http://dx.doi.org/10.1007/JHEP08(2011)054}{{\em JHEP} {\bfseries 08}
  (2011) 054}, \href{http://arxiv.org/abs/1105.4628}{{\ttfamily arXiv:1105.4628
  [hep-ph]}}. [Erratum: JHEP 10, 101 (2017)].

\bibitem{Kelley:2011ng}
R.~Kelley, M.~D. Schwartz, R.~M. Schabinger, and H.~X. Zhu, ``{The two-loop
  hemisphere soft function},''
  \href{http://dx.doi.org/10.1103/PhysRevD.84.045022}{{\em Phys. Rev. D}
  {\bfseries 84} (2011) 045022},
  \href{http://arxiv.org/abs/1105.3676}{{\ttfamily arXiv:1105.3676 [hep-ph]}}.

\bibitem{Schwartz:2014wha}
M.~D. Schwartz and H.~X. Zhu, ``{Nonglobal logarithms at three loops, four
  loops, five loops, and beyond},''
  \href{http://dx.doi.org/10.1103/PhysRevD.90.065004}{{\em Phys. Rev.}
  {\bfseries D90} no.~6, (2014) 065004},
\href{http://arxiv.org/abs/1403.4949}{{\ttfamily arXiv:1403.4949 [hep-ph]}}.

\bibitem{Caron-Huot:2015bja}
S.~Caron-Huot, ``{Resummation of non-global logarithms and the BFKL
  equation},'' \href{http://dx.doi.org/10.1007/JHEP03(2018)036}{{\em JHEP}
  {\bfseries 03} (2018) 036}, \href{http://arxiv.org/abs/1501.03754}{{\ttfamily
  arXiv:1501.03754 [hep-ph]}}.

\bibitem{Hagiwara:2015bia}
Y.~Hagiwara, Y.~Hatta, and T.~Ueda, ``{Hemisphere jet mass distribution at
  finite $N_c$},'' \href{http://dx.doi.org/10.1016/j.physletb.2016.03.028}{{\em
  Phys. Lett. B} {\bfseries 756} (2016) 254--258},
  \href{http://arxiv.org/abs/1507.07641}{{\ttfamily arXiv:1507.07641
  [hep-ph]}}.

\bibitem{Larkoski:2015zka}
A.~J. Larkoski, I.~Moult, and D.~Neill, ``{Non-Global Logarithms,
  Factorization, and the Soft Substructure of Jets},''
  \href{http://dx.doi.org/10.1007/JHEP09(2015)143}{{\em JHEP} {\bfseries 09}
  (2015) 143},
\href{http://arxiv.org/abs/1501.04596}{{\ttfamily arXiv:1501.04596 [hep-ph]}}.

\bibitem{Becher:2015hka}
T.~Becher, M.~Neubert, L.~Rothen, and D.~Y. Shao, ``{Effective Field Theory for
  Jet Processes},''
  \href{http://dx.doi.org/10.1103/PhysRevLett.116.192001}{{\em Phys. Rev.
  Lett.} {\bfseries 116} no.~19, (2016) 192001},
\href{http://arxiv.org/abs/1508.06645}{{\ttfamily arXiv:1508.06645 [hep-ph]}}.

\bibitem{Balsiger:2018ezi}
M.~Balsiger, T.~Becher, and D.~Y. Shao, ``{Non-global logarithms in jet and
  isolation cone cross sections},''
  \href{http://dx.doi.org/10.1007/JHEP08(2018)104}{{\em JHEP} {\bfseries 08}
  (2018) 104}, \href{http://arxiv.org/abs/1803.07045}{{\ttfamily
  arXiv:1803.07045 [hep-ph]}}.

\bibitem{Banfi:2021owj}
A.~Banfi, F.~A. Dreyer, and P.~F. Monni, ``{Next-to-leading non-global
  logarithms in QCD},'' \href{http://arxiv.org/abs/2104.06416}{{\ttfamily
  arXiv:2104.06416 [hep-ph]}}.

\bibitem{Appleby:2002ke}
R.~Appleby and M.~Seymour, ``{Nonglobal logarithms in interjet energy flow with
  kt clustering requirement},''
  \href{http://dx.doi.org/10.1088/1126-6708/2002/12/063}{{\em JHEP} {\bfseries
  12} (2002) 063}, \href{http://arxiv.org/abs/hep-ph/0211426}{{\ttfamily
  arXiv:hep-ph/0211426}}.

\bibitem{Delenda:2006nf}
Y.~Delenda, R.~Appleby, M.~Dasgupta, and A.~Banfi, ``{On QCD resummation with
  $k_t$ clustering},''
  \href{http://dx.doi.org/10.1088/1126-6708/2006/12/044}{{\em JHEP} {\bfseries
  12} (2006) 044}, \href{http://arxiv.org/abs/hep-ph/0610242}{{\ttfamily
  arXiv:hep-ph/0610242}}.

\bibitem{Neill:2018yet}
D.~Neill, ``{Non-Global and Clustering Effects for Groomed Multi-Prong Jet
  Shapes},'' \href{http://dx.doi.org/10.1007/JHEP02(2019)114}{{\em JHEP}
  {\bfseries 02} (2019) 114}, \href{http://arxiv.org/abs/1808.04897}{{\ttfamily
  arXiv:1808.04897 [hep-ph]}}.

\bibitem{Cacciari:2008gp}
M.~Cacciari, G.~P. Salam, and G.~Soyez, ``{The anti-k$_T$ jet clustering
  algorithm},'' \href{http://dx.doi.org/10.1088/1126-6708/2008/04/063}{{\em
  JHEP} {\bfseries 04} (2008) 063},
\href{http://arxiv.org/abs/0802.1189}{{\ttfamily arXiv:0802.1189 [hep-ph]}}.

\bibitem{Dulat:2015mca}
S.~Dulat, T.-J. Hou, J.~Gao, M.~Guzzi, J.~Huston, P.~Nadolsky, J.~Pumplin,
  C.~Schmidt, D.~Stump, and C.~P. Yuan, ``{New parton distribution functions
  from a global analysis of quantum chromodynamics},''
  \href{http://dx.doi.org/10.1103/PhysRevD.93.033006}{{\em Phys. Rev.}
  {\bfseries D93} (2016) 033006},
\href{http://arxiv.org/abs/1506.07443}{{\ttfamily arXiv:1506.07443 [hep-ph]}}.

\bibitem{Ligeti:2008ac}
Z.~Ligeti, I.~W. Stewart, and F.~J. Tackmann, ``{Treating the b quark
  distribution function with reliable uncertainties},''
  \href{http://dx.doi.org/10.1103/PhysRevD.78.114014}{{\em Phys. Rev.}
  {\bfseries D78} (2008) 114014},
\href{http://arxiv.org/abs/0807.1926}{{\ttfamily arXiv:0807.1926 [hep-ph]}}.

\bibitem{Note1}
The ALICE normalization is not normalized by a few percent, because of their
  treatment of jets that never pass the soft drop condition.

\bibitem{Arratia:2019vju}
M.~Arratia, Y.~Song, F.~Ringer, and B.~V. Jacak, ``{Jets as precision probes in
  electron-nucleus collisions at the future Electron-Ion Collider},''
  \href{http://dx.doi.org/10.1103/PhysRevC.101.065204}{{\em Phys. Rev. C}
  {\bfseries 101} no.~6, (2020) 065204},
  \href{http://arxiv.org/abs/1912.05931}{{\ttfamily arXiv:1912.05931
  [nucl-ex]}}.

\bibitem{Page:2019gbf}
B.~S. Page, X.~Chu, and E.~C. Aschenauer, ``{Experimental Aspects of Jet
  Physics at a Future EIC},''
  \href{http://dx.doi.org/10.1103/PhysRevD.101.072003}{{\em Phys. Rev. D}
  {\bfseries 101} no.~7, (2020) 072003},
  \href{http://arxiv.org/abs/1911.00657}{{\ttfamily arXiv:1911.00657
  [hep-ph]}}.

\end{thebibliography}\endgroup

\end{document}